\documentclass[conference,compsoc]{IEEEtran}

\usepackage{cite}
\usepackage{subcaption}
\usepackage{color}

\usepackage{listings}
\definecolor{codebackground}{rgb}{0.97,0.97,0.87}
\usepackage{graphicx}
\DeclareGraphicsExtensions{.pdf,.jpeg,.png}
\usepackage{amsmath}

\begin{document}
\title{Performance Models for Split-execution Computing Systems}
\author{
\IEEEauthorblockN{Travis S. Humble\IEEEauthorrefmark{1}\IEEEauthorrefmark{2},
Alexander J. McCaskey\IEEEauthorrefmark{1},
Jonathan Schrock\IEEEauthorrefmark{1},\\ 
Hadayat Seddiqi\IEEEauthorrefmark{1},
Keith A. Britt\IEEEauthorrefmark{1}\IEEEauthorrefmark{2} and
Neena Imam\IEEEauthorrefmark{1}\IEEEauthorrefmark{3}}
\IEEEauthorblockA{\IEEEauthorrefmark{1}Quantum Computing Institute, Oak Ridge National Laboratory, Oak Ridge, Tennessee, 37831}
\IEEEauthorblockA{\IEEEauthorrefmark{2}Bredesen Center, University of Tennessee, Knoxville, Tennessee, 37996}
\IEEEauthorblockA{\IEEEauthorrefmark{3}Computational Research and Development Programs, Oak Ridge National Laboratory, Oak Ridge, Tennessee, 37831}
}

\maketitle

\begin{abstract}
Split-execution computing leverages the capabilities of multiple computational models to solve problems, but splitting program execution across different computational models incurs costs associated with the translation between domains. We analyze the performance of a split-execution computing system developed from conventional and quantum processing units (QPUs) by using behavioral models that track resource usage. We focus on asymmetric processing models built using conventional CPUs and a family of special-purpose QPUs that employ quantum computing principles. Our performance models account for the translation of a classical optimization problem into the physical representation required by the quantum processor while also accounting for hardware limitations and conventional processor speed and memory. We conclude that the bottleneck in this split-execution computing system lies at the quantum-classical interface and that the primary time cost is independent of quantum processor behavior.
\end{abstract}

\IEEEpeerreviewmaketitle

\section{Introduction}
The discovery of quantum algorithms showing exponential speed ups in the limit of large problem sizes suggests that this computational model may have a profound impact on the performance of future computing systems \cite{Nielsen2000}. This has led to the fast-paced development of devices that implement the principles of quantum computing and can ultimately test algorithmic implementations \cite{Ladd2010}. Despite theoretical expectations for quantum algorithms, the actual computational performance expected from quantum processing units (QPUs) remains unclear. This is due, in part, to the poorly explored interface between QPUs and conventional computing systems. In addition, much of quantum computing has remained a theoretical exercise supported by modest experiments to demonstrate proof of principle behavior as opposed to practical performance. An important point not examined by current experimental studies is that a computing system comprising both quantum and conventional models must support a split-execution environment. This means that the system and the application operate within the context of two distinct computational paradigms, e.g., quantum and classical. 
\par
Recently, a complete QPU has been realized as a special-purpose variant of a quantum computing system. The family of processors from D-Wave Systems, Inc. has been used for remarkable demonstrations of how a different computational paradigm can correctly solve difficult mathematical problems using the principles of quantum computing \cite{Harris2010,Johnson2011}. The D-Wave QPUs are special-purpose processors in the sense that they solve a specific discrete optimization problems, and they use only a limited subset of quantum computing principles. Notwithstanding these restrictions, a variety of interesting problems have been shown to map into the D-Wave processor, including including classification \cite{Neven2008a,Neven2008b}, machine learning \cite{Pudenz2013}, graph theory \cite{Gaitan2012,Hen2012,Bian2013,Gaitan2014}, and protein folding \cite{PerdomoOrtiz2008,PerdomoOrtiz2012} among others \cite{Smelyanskiy2012,Lucas2014,Vinci2014,PerdomoOrtiz2014a,OGorman2014,Rieffel2015}. In addition, genuine quantum behavior has been observed in these devices \cite{Ronnow2014,Lanting2014}.
\par
The availability of the D-Wave QPU has forced the question of how quantum computing devices can be used within actual computing systems. Although performance of the existing processor has been analyzed with respect to isolated computational time \cite{McGeoch2013,Rieffel2015,King2015}, the overall performance of the processor integrated into a larger host system has not yet been examined. In particular, the relative computational complexities facing the translation between computational domains have not been compared, and there is good reason to believe that these steps may represent significant hurdles to practical performance gains. These complexities of translating data need not exhibit the same performance as the underlying algorithm and, in the limit of large input data, domain translation poses the potential to remove any advantage offered by algorithm performance within a different computational model.
\par
\begin{figure}
\textbf{(a)}\hspace{0cm}
\begin{subfigure}{0.5\textwidth}
\centerline{\includegraphics[height=1.5cm]{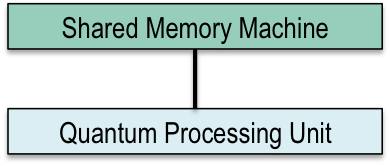}}
\end{subfigure}
\textbf{(b)}\hspace{0cm}
\begin{subfigure}{0.5\textwidth}\vspace{1em}
\centerline{\includegraphics[height=2cm]{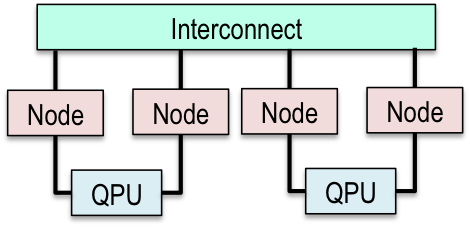}}
\end{subfigure}
\textbf{(c)}\hspace{0cm}
\begin{subfigure}{0.5\textwidth}\vspace{1em}
\centerline{\includegraphics[height=2.1cm]{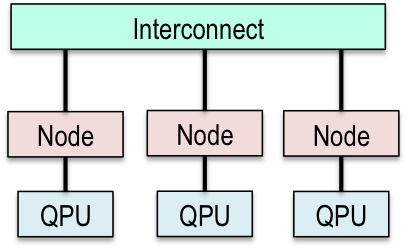}}
\end{subfigure}
\caption{Three architectural models for integrating a QPU into a host HPC system: (a) asymmetric multi-processor, (b) shared-resource, and (c) a dedicated QPU for each node.}
\label{fig:split_execution_arch}
\end{figure}

As shown in Fig.~\ref{fig:split_execution_arch}, there are several architectures amenable to split-execution computing differentiated by how the host system supports the QPU. A comparision across all three architectures has been recently made by Britt and Humble \cite{Britt2015}, and we will limit the subsequent discussion to the asymmetric multi-processor model represented by Fig.~\ref{fig:split_execution_arch}(a). Our choice is motivated by current infrastructure constraints on the existing D-Wave QPU that prevent it from being more tightly integrated. Consequently, the loosely integrated architecture in Fig.~\ref{fig:split_execution_arch}(a) is indicative of our near-term expectations for coupling a D-Wave quantum processor with existing conventional computing systems. This loose architecture may be understood conceptually as a classical client requesting a response from a quantum server via a local area network interface.
\par
In this contribution, we examine how the interaction between the two computing systems influences the overall run-time performance. We focus on the need to translate between the classical and quantum computational domains. This is a unique aspect of split-execution computing systems that we find to contribute a non-trivial cost to the overall run-time behavior. In order to explore this question, we have developed a performance model for the asymmetric system that addresses translating program statements from a CPU to QPU and back again. Our implementation builds on previous efforts to develop software simulators for quantum devices \cite{Humble2014}, and the current work extends these ideas to the performance of integrated systems by tracking the behavior of both quantum and classical elements. Our approach provides timing estimates for the different stages of the split-execution system that are considered to be representative of the application-level behavior that arises using the D-Wave family of processors. Our analysis emphasizes how the quantum and classical paradigms interact and provides insights into the overall time-to-solution needed for solving a general application within this programming model.
\par
The remainder of the presentation is organized as follows. Sec.~\ref{sec:sem} presents a definition of the split-execution model including specifications about the functionality of the QPU. Sec.~\ref{sec:performance} describes our performance model for the combined system, which is based on the ASPEN hardware modeling language, and includes description of a generic application for the discrete optimization solver. We also present the results from using our models to capture timing estimates at each stage of the split-execution model, while Sec.~\ref{sec:con} offers conclusions on these results.
\section{Split-execution Computing System}
\label{sec:sem}
\begin{figure}
\centering
\includegraphics[width=.95\columnwidth]{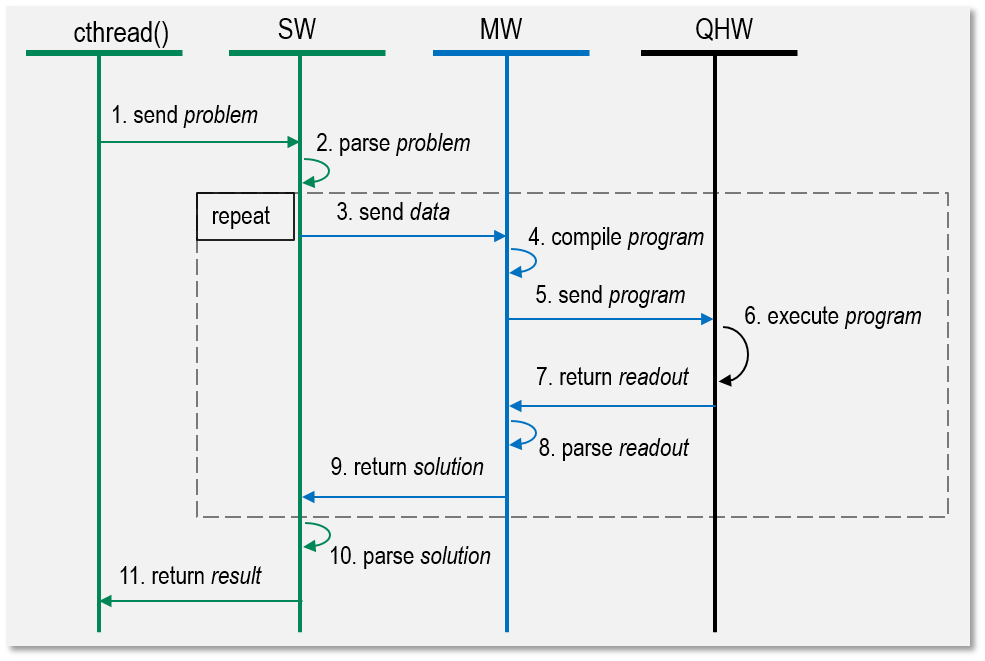}
\caption{A sequence diagram describing how a CPU interacts with a QPU. The calling thread executes in a conventional computing environment, while the software (SW) and middleware (MW) perform domain translation to the quantum computing environment. The latter executes on quantum hardware (QHW) before returning a result that must be translated back to the conventional computing model.}
\label{fig:seq_diag}
\end{figure}

A simple example of how a host system drives QPU execution is presented in Fig.~\ref{fig:seq_diag}. In this example, a calling thread \emph{cthread} running on the host CPU pushes the data defining an input \emph{problem} to the QPU. The interface to the QPU is defined by a software (SW) layer that parses the incoming \emph{problem} into the data needed for driving the device middleware (MW) layer. The MW layer is responsible for constructing from the data a \emph{program} that represents the sequence of operations to be performed by the quantum hardware (QHW). The \emph{program} is executed by the QHW layer, e.g., through the action of an electronic control system and associated signal generators. The \emph{readout} data obtained after QHW execution corresponds with measurement of the quantum system, which effectively generates a classical representation of the quantum computation. The \emph{readout} is returned to the MW layer, where it may undergo additional post-processing to construct a \emph{solution} to the original \emph{problem} before returning to the SW layer. These latter stages may be repeated multiple times, for example, to collect statistical samples from the execution of a program expected to be behave probabilistically \cite{Humble2014}.
\par
The run-time example in Fig.~\ref{fig:seq_diag} neglects many of the problems facing a realistic computing scenario, including resource competition, caching, concurrency, etc., but it does highlight the fundamental challenge involved in the use of a QPU. Namely, the context in which computation occurs changes as the data moves from the SW layer to the QHW layer. The MW layer is responsible for negotiating this change within Fig.~\ref{fig:seq_diag}. The specific transformation imposed by the MW will depend on the underlying computational models, and the ease with which these transformations may be implemented will vary. Our analysis will focus on the the case of transforming between a conventional random-access memory (RAM) model and the adiabatic quantum computing (AQC) model that has been partially realized by the D-Wave QPU.
\subsection{The D-Wave QPU}
As an example of split-execution computing, we develop a model for an asymmetric multi-processor system that includes a QPU based on the family of processors from D-Wave Systems \cite{Johnson2010,Johnson2011}.  We briefly summarize the programming model for the D-Wave QPU, which has been discussed extensively elsewhere \cite{Harris2010,McGeoch2013,Humble2014,Rieffel2015,King2015,Britt2015}. 
\par
The basic principle of adiabatic quantum computing is to leverage continuous-time quantum dynamics that adiabatically transform an initial quantum register state into a final computational result \cite{Kadowaki1998,Farhi2000, Farhi2000b}. This is represented formally by the Schrodinger equation,
\begin{equation}
i\frac{\partial \Psi(t)}{\partial t} = H(t) \Psi(t),
\end{equation}
where $\Psi(t)$ is the quantum state of the register at time $t$ and $H(t)$ is the Hamiltonian that defines the interactions between register elements. The role of the dynamics is to affect computation on the register and change its value, namely, $\Psi(t)$. Conceptually, this occurs by adiabatically deforming the Hamiltonian that governs the underlying interactions betwen register elements. For a register initially prepared in the lowest energy state of the starting Hamiltonian, i.e., the ground state, adiabatic dynamics ensure the register remains a ground state of the instantaneous Hamiltonian throughout the computation. This basic principle is used to recover the ground state of a known Hamiltonian. Readout then corresponds with collapsing the register elements into definite classical values that provide a computational solution. 
\par
\begin{figure}
\begin{minipage}{0.5\textwidth}
\centerline{\includegraphics[width=0.85\columnwidth]{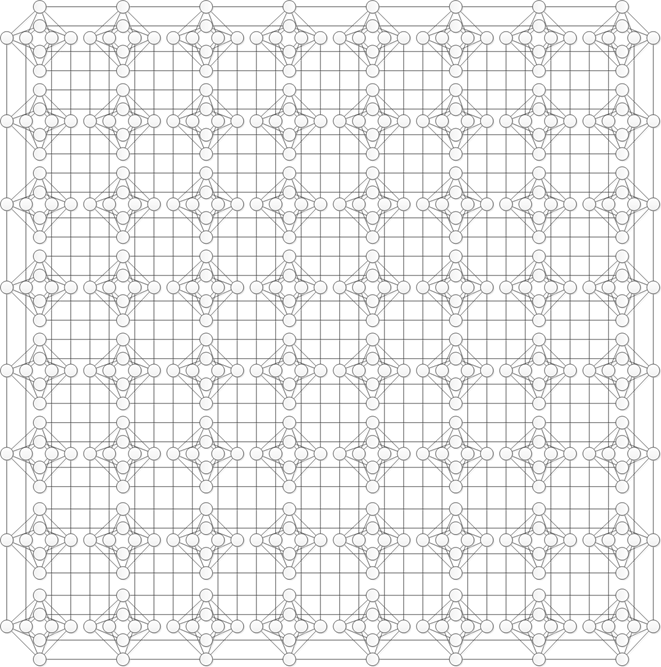}}
\vspace{1em}
\end{minipage}
\caption{The hardware connectivity graph describing the D-Wave processor family, in which nodes represent physical qubits and edges signify tunable interactions. This diagram presents 512 qubits, or an 8-by-8 lattice of unit cells. The most recent processor supports a 12-by-12 lattice and 1152 qubits.}
\label{fig:chimera}
\end{figure}

The D-Wave QPU implements a restricted form of the adiabatic quantum computing model that limits the range of computations it realizes. These restrictions are enforced by the processor hardware, which is based on an array of interacting superconducting Josephson junctions laid out in a square grid of bipartite graphs know as a Chimera graph as shown in Fig.~\ref{fig:chimera}. For the D-Wave QPU, the Chimera hardware layout restricts each qubit to interact with 6 neighbors (5 neighbors in the case of edge qubits). This limited connectivity is one restriction on the computational model, while another is that the programmable interactions between qubits must adhere to a Hamiltonian of Ising form, i.e., 
\begin{equation}
\label{eq:h0}
H_{\textrm{Ising}} = -\sum_{i=1}^{n}{h_{i} \hat{Z}_{i}} - \sum_{i,j=1}^{n}{\hat{J}_{i,j} \hat{Z}_{i} \hat{Z}_{j}},
\end{equation}
where $n$ is the number of qubits, $h_i$ is a real-valued bias on qubit $i$, $J_{i,j}$ is a real-valued coupling between qubits $i$ and $j$, and $\hat{Z}_{i}$ is the Pauli operator \cite{Nielsen2000}. In order to completely realize the adiabatic quantum computing mode, the Hamiltonian would be required to support additional interaction terms \cite{Biamonte2008}.
\par
The limitation to only $\hat{Z}\hat{Z}$ interactions in Eq.~(\ref{eq:h0}) prevents the D-Wave QPU from implementing the full range of adiabatic quantum computing. However, despite this restriction on the Hamiltonian, there is a wide variety of problems that fall within the purview of the current restrictions. A definitive example is given in the case of a quadratic unconstrained binary optimization (QUBO) problem \cite{Boros2002}, i.e., finding
\begin{equation}
\label{eq:qubo}
\arg \min_{b} b^{T} Q b,
\end{equation}
where $b$ is a binary vector of length $n$ and $Q$ is an $n$-by-$n$ symmetric real-valued matrix. Solving the QUBO problem is generally NP-hard, and therefore, it is unlikely that efficient algorithms for general instances exist even within models of quantum computation. However, the D-Wave processors permit the exploration of new efficient heuristics for solving QUBO problems and other related problems. These chiefly include optimization problems or reductions to optimization problems such as \textsc{max-sat}, \textsc{min-cover}, \textsc{max-cut} and other graph problems as well as binary classification, integer linear programming, and set packing problems among many others \cite{Lucas2014}. In each case, the common QUBO form can be reduced to the Ising Hamiltonian in Eq.~(\ref{eq:h0}) by a straightforward mapping of one quadratic form into another \cite{Choi2008,Choi2011,Klymko2014,Humble2014}. We discuss these programming aspects in more detail below.
\begin{figure}
\begin{minipage}{0.5\textwidth}
\centerline{\includegraphics[width=0.65\columnwidth]{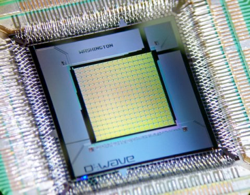} }
\end{minipage}
\caption{A photograph of the latest member of the DW processor family: the DW2X processor is composed from 1152 physical qubits, whose connectivity graph is shown in Fig.~\ref{fig:chimera}. Credit: D-Wave Systems, Inc.}
\label{fig:dw2x}
\end{figure}
\par
The latest member in the D-Wave family is a 1152-qubit processor released in 2015 \cite{King2015}. Instances of this design have been installed in several third-party locations including NASA Ames Research Center and University of Southern California Information Sciences Institute. In addition to the processor, management and control electronics as well as infrastructure to shield and cool the processor are required for QPU operation. Cooling requires a dilution refrigerator, which is a large, bulky and expensive system capable of lowering the operating temperature of the processor to 14 mK. An example of the QPU is shown in Fig.~\ref{fig:dw2x}.
\subsection{Programming the D-Wave QPU}
Programming the D-Wave QPU requires translation of an input problem, such as a QUBO or Ising instance, into the physical Ising Hamiltonian form that satisfies the hardware layout constraints \cite{Humble2014}. An intermediate form of this mapping will first construct a logical Ising model of the QUBO instance in which the hardware constraints are ignored, e.g., connectivity. We refer to this intermediate data structure as the logical Ising model, because it represents an abstraction of the QUBO in terms of an idealized logical spin system. Construction of the logical Ising model from the matrix $Q$ defined in Eq.~(\ref{eq:qubo}) is a relatively simple process \cite{Humble2014}, i.e., Ising parameters are defined as
\begin{equation}
\label{eq:hi}
h_{i} = \frac{1}{2} Q_{i,i} + \frac{1}{4}\sum_{j=1}^{n}{Q_{i,j}}\hspace{1cm}\textrm{for }i = 1\textrm{ to }n
\end{equation}
and
\begin{equation}
\label{eq:jij}
J_{i,j} = \frac{1}{4} Q_{i,j}\hspace{1cm}\textrm{for }i < j = 1\textrm{ to }n.
\end{equation}
In total, these steps require $O(n^3)$ addition operations. 
\par
By contrast, mapping the logical Ising model to the constrained Ising hardware is a more computationally difficult problem. Formally, this process is cast as a minor graph embedding problem, in which the connectivity graph for the logical Ising model must be embedded into the physical hardware graph \cite{Choi2008,Klymko2014}. Let $G$ denote the input graph that has the (weighted) adjacency matrix $Q$ and let $H$ denote the graph expressing the underlying hardware connectivity. A minor embedding of $G$ into $H$ is defined as a mapping $\phi$ of each vertex in $G$ into $H$ such that 1)  every vertex in $G$ maps to a connected subtree of $H$, and 2) every edge in $G$ maps to an edge between corresponding subtrees in $H$. We will denote the corresponding graph as $\phi(G)$ and we will call $G$ embeddable in $H$ if the mapping exists. Note that the graph $\phi(G)$ can be substantially larger than the original graph $G$. This is the result of constructing large subtrees within $H$ to ensure the required edges in $\phi(G)$ are realized. 
\par
For a general input graph $G$ and arbitrary hardware graph $H$, the problem of constructing $\phi(G)$ is an FNP-complete problem, the functional variant of an NP-complete problem. Previous work has identified polynomial heuristics for the case of embedding complete input graphs into the Chimera hardware graphs employed in the D-Wave QPU \cite{Choi2008,Klymko2014}. For these methods the embedding of an input graph with $n$ vertices requires a Chimera hardware with $n^2$ qubits. While this is necessary for embedding a complete graph, input problems are not necessarily fully connected and the same methods will overestimate the number of hardware qubits required for an embedding and thus limit the size of problems that can be solved. Consequently, there is a need to develop methods of embedding into Chimera graphs that use fewer qubits by taking into account properties of the input graph.
\par
Alternatives include a brute force approach to minor embedding that relies on solving the subgraph isomorphism problem to identify the smallest embedded minor. Although that approach has a computational complexity that scales exponentially with hardware size, it may be possible to precompute these maps offline and store them as lookup tables. A second alternative is to use a non-deterministic technique recently proposed by Cai, Macready, and Roy to reduce the average number of hardware qubits used \cite{Cai2014}. Their approach employs Djikstra's  algorithm to construct the minimum path between randomly distributed subtrees. Although the worst-case complexity for this algorithm grows as $O(n^3)$, we will use this method in developing our programming resource model below as it permits the largest sized input problems to be programmed on the processor. Moreover, the average case complexity was observed by Cai et al.~to be significantly less, i.e., $O(n)$ \cite{Cai2014}. 
An important point for all of the aforementioned algorithms is that they must take into account the random faults in the hardware graph that can occur during fabrication. These faulty qubits and couplers are readily identified during processor calibration and must be deactivated to avoid unwanted usage. The loss of a node within the Chimera layout can destroys its underlying symmetry and, consequently, make the embedding problem more difficult \cite{Klymko2014}.
\par
After finding a minor embedding of the logical Ising model into the hardware, the corresponding parameters for the embedded Ising model must be set. This step is relatively straightforward given the Ising parameters defined by Eqs.~(\ref{eq:hi}) and (\ref{eq:jij}). However, the ability to realize these exact parameter values is limited by the bits of precision expressed by the electronic control system and the hardware couplers. This implies that the final, programmed Ising model may be substantively different from the intended logical input. It is not yet clear what errors these differences contribute to final solutions of the QUBO problem or what methods may mitigate these errors. Finally, one additional coupling strength must be introduced to account for the interactions between qubits forming embedded subtrees within the hardware. In practice, this value is typically chosen to be much larger than neighboring elements to ensure all qubits within a subgraph behave collectively \cite{Choi2008}.
\par
Other programming  choices include the schedule for annealing the system to the final Hamiltonian, e.g., characterized by the temporal waveform and duration. Limitations on the hardware control system do not allow for arbitrary waveforms and duration but restrict these options to pre-defined ranges. Execution of the program within the hardware also requires pre-processing steps to initialize the electronic control system and construct the analog signals applied to the quantum chip. These steps include the time required to initialize the programmable magnetic memory (PMM) that is used as the control lines into the super-cooled processor. Technical details about the electronic programming process are available in the relevant literature \cite{Johnson2010}, but it suffices to note that these steps contribute a near constant time cost to the total for the execution model.
\section{Split-execution Performance Model}
\label{sec:performance}
The behavior of adiabatic quantum dynamics under the Hamiltonian in Eq.~(\ref{eq:h0}) is a subject of intense scrutiny for physicists interested in quantum physical properties and now also computer scientists. The relevant question for these two communities are related but nonetheless different. For purposes of performing computation, the performance and behavior of a QPU relative to other aspects of a computing system are essential to understanding its ultimate utility. Recent experiments investigating the time-to-solution for using a D-Wave QPU offer insights into its isolated behavior \cite{McGeoch2013,Rieffel2015,King2015}, but the question of dependency with other computing elements requires more elaborate models of the system. In this section, we present models for the interaction between a D-Wave QPU and a host CPU that highlight the dependency on data movement and the critical need to convert the classical program into a quantum program representation. 
\subsection{Machine Models}
We have developed a model for the asymmetric design that includes both classical and quantum hardware using ASPEN. ASPEN is a hardware modeling language developed at ORNL for structured analytical performance modeling \cite{Spafford2012, Spafford2015,ASPEN}. It is a formal language that can be used to generate representations of applications and abstract machine models and analyze the predicted performance of these models. ASPEN was used to generate representations of how QPU's and quantum-enabled applications will perform. We have developed machine models that account for a single node composed from a CPU and a QPU. For the CPU, we have used existing ASPEN models based on the Intel Xeon processor while for the QPU we have developed a model specific to the D-Wave family of quantum processors. We used the above description of the D-Wave processor programming model to develop a hardware model of the processor in the ASPEN language.  
\begin{figure}
\begin{lstlisting}
include memory/ddr3_1066.aspen
include sockets/intel_xeon_e5_2680.aspen
include sockets/nvidia_m2090.aspen
include sockets/dwave_vesuvius_20.aspen

machine SimpleNode 
{
  [1] SIMPLE nodes
}

node SIMPLE 
{
   [1] intel_xeon_e5_2680 sockets
   [1] nvidia_m2090 sockets 
   [1] DwaveVesuvius20 sockets 
}
\end{lstlisting}
\begin{lstlisting}
socket DwaveVesuvius {
   [1] Vesuvius cores 

   gddr5 memory
   linked with pcie
}

core Vesuvius20 {
     resource QuOps(number) [number * 20/1000000]
}
\end{lstlisting}
\caption{The ASPEN model for the CPU-QPU node and the D-Wave Vesuvius hardware socket}
\label{fig:aspen_hardware}
\end{figure}
\par
An example of the ASPEN model for the host system node is shown in Fig.~\ref{fig:aspen_hardware}. 
This represents a \emph{SimpleNode} that consists of an Intel Xeon CPU, an NVidia GPU, and a D-Wave 
Vesuvius QPU. These are each expressed as sockets within the machine model. The corresponding 
model for the QPU socket is shown in Fig.~\ref{fig:aspen_hardware}. The model for the socket 
declares how many cores the QPU has and, due to the ASPEN syntax, must also declare a classical 
memory element. We do not however make use of these memory elements in our analysis. Also note 
that the socket model includes a PCIe interface, which is used for the movement of data between 
the CPU and QPU. The D-Wave QPUs currently support a client-server interface but we do not model 
those aspects of the interface in this work. As shown below, networking is not expected to be 
the dominant cost of hardware model. Figure~\ref{fig:aspen_hardware} also models the core of 
the QPU. This model converts the number of quantum operation (\texttt{QuOps}) resources used by the core to an execution time. In the model shown in Fig.~\ref{fig:aspen_hardware}, an annealing duration of 20 $\mu s$ is shown but as noted previously this duration may be scaled according to program options (the D-Wave QPUs currently set the annealing duration to 20 $\mu s$ in in the absence of user input).
\subsection{Application Models}
The machine model is distinct from the application model that drives program execution. The 
latter depends on the implementation details of the application, including the calculations needed 
to pre-process and post-process data for the quantum program as well as any quantum calculations. 
There are a variety of problems that can be mapped into the Ising Hamiltonian that underlies 
the D-Wave processor, each with a distinct set of stages needed to manage interactions with 
the QPU. In this work, we investigate the generic problem of embedding an Ising Hamiltonian into the D-Wave QPU, finding the lowest energy state of that Hamiltonian, and converting the resulting readout to a solution.
\par
We decompose the application model into three stages. Stage 1 denotes the pre-processing needed to embed the logical Hamiltonian into the hardware graph. It also includes the steps needed to subsequently set the hardware parameters and initialize the electronic control structures used during program execution. An ASPEN model for the first stage is shown in Fig.~\ref{fig:stage1}, in which \texttt{LPS} is the logical problem size defining the number of spins (or vertices) in the logical Hamiltonian. In the listing of Fig.~\ref{fig:stage1}, we have used the worst-case estimate of the embedding complexity (number of operations) taken from Cai et al. \cite{Cai2014}, which uses a non-deterministic search for finding the embedding in a Chimera hardware. We further assume that that size of the embedded graph is the largest possible embedded graph, i.e., $\texttt{LPS}^2$. Therefore, the resulting ASPEN model for stage 1 includes a series of resource consumption statements to account for the calculation of the embedding. We also include stage 1 the constant steps required to initialize the electronic control system that drives the quantum hardware layer. In this case, we have used times shown in microseconds that reflect the average durations required for performing the initializations within the second generation, DW2 Vesuvius processor, and we assume these constants are nearly the same within the latest DW2X processor from D-Wave.
\begin{figure}
\begin{lstlisting}
model Stage1
{
  param LPS                     = 0 // Input Parameter
  param Ising                   = LPS^2 
  param NH = LPS 
  param EH = NH*(NH-1) / 2
  param M = 12
  param N = 12
  param NG = 8*M*N 
  param EG = 4*(2*M*N - M - N) + 16*M*N 
  param EmbeddingOps = (EG+NG*log(NG))*(2*EH)*NH*NG   
  param ParameterSetting = LPS^ 3 

  // Hardware constants for DW2 in microseconds
  param StateCon       = 252162 
  param PMMSW        = 33095
  param PMMElec      = 0    
  param PMMChip     = 11264
  param PMMTherm  = 10000 
  param SWRun         = 4000 
  param ElecRun        = 9052 
  param ProcessorInitialize = StateCon+PMMSW+PMMElec+PMMChip+PMMTherm+SWRun+ElecRun

  data Input as Array((NH*NH), 4)
  data Output as Array((NG* NG), 4)

  kernel InitializeData  {
    execute [1] {
      flops [Ising] as sp, fmad, simd
      stores [NH*4] to Input
    }

    execute [1] {
      flops [ParameterSetting] as sp, fmad, simd
    }
  }

  kernel EmbedData  {
    execute embed [1] {
      loads [EH*4] from Input
      flops [EmbeddingOps] as sp, simd
      stores [EG*4] to Output
      intracomm [EG*4] as copyout
    }
  }

  kernel InitializeProcessor  {
    execute [1] {microseconds [ProcessorInitialize]}
  }

  kernel main
  {
    InitializeData
    EmbedData
    InitializeProcessor
  }
}
\end{lstlisting}
\caption{Stage 1 of the split-execution application implements the generation and embedding of a logical Ising Hamiltonian into the D-Wave processor.}
\label{fig:stage1}
\end{figure}
\par
The second stage of the application model captures the behavior of the adiabatic quantum dynamics performed by the D-Wave QPU. This model includes the time required to initialize the processor register and to readout the register as well as the number of \texttt{QuOps} incurred during program execution. In particular, we reduce \texttt{QuOps} for the D-Wave processor into a time cost by accounting for how many repetitions of the annealing sequence are performed during a single call to the QPU. Because the D-Wave QPU is effectively a probabilistic processor, multiple runs are required to collect statistics and build confidence that the lowest observed energy state is likely the global minimum. In the ASPEN model shown in Fig.~\ref{fig:stage2}, we set the number of required repetitions based on the desired solution accuracy $p_a$ and the characteristic probability $p_s$ that any single run finds the lowest-energy state. The required number of iterations $s$ is given as 
\begin{equation}
s \geq \frac{\log(1-p_a)}{\log(1-p_s)},
\end{equation}
which sets the number of \texttt{QuOps} needed.
\begin{figure}
\begin{lstlisting}
model Stage2
{
  param Success                 = 0.9999  
  param Accuracy                = 0 // Input parameter
  param AnnealReadResults       = 320  
  param AnnealThermalization    = 5     

  kernel Stage2Processing
  {
    execute mainblock2[1]
    {
        // Number of QPU calls
        QuOps [ceil(log(1-(Accuracy/100))/log(1-Success))]
    }
    execute mainblock3[1]
    {
        // Readout time
        microseconds [AnnealReadResults]
    }
    execute mainblock4[1] {
        // Initialization time
        microseconds [AnnealThermalization]
    }
  }

  kernel main  {
     Stage2Processing
  }
}
\end{lstlisting}
\caption{Stage 2 of the split-execution application uses the D-Wave processor as an optimization solver that performs statistical sampling to recover the lowest energy state.}
\label{fig:stage2}
\end{figure}
\par
At this point, we should note that the probability $p_s$ for the register to be in the ground state is strongly influenced by whether the evolution was sufficiently slow relative to the internal 
dynamical timescales. In particular, it is known that the leading order error in adiabatic quantum computing arises from excitations to higher-lying energy eigenstates, where the transition probability is determined by the minimal instantaneous energy gap between the ground and (first) excited manifolds \cite{Farhi2000}. The time complexity of this probabilistic algorithm is thus determined by the shortest evolutionary period $T$ needed to obtain a probability $p_e$ to be in the ground state. This value is not arbitrary as success depends on the annealing 
time $T$ and the shape of the annealing schedule as well as the internal energy structure of the Ising Hamiltonian. However, as we will emphasis from our results, the exact value of the characteristic success has a relatively small influence of the overall performance of the integrated application model. 
\par
The third and final stage in our application model defines the post-processing behavior. This includes sorting the ensemble of readout results gathered during multiple runs of the programs as well as the post-processing required by the QPU to return these results to the CPU. For the sort, we assume an underlying heapsort algorithm is used to sort the readout results according to the value of the computed energy. Although only the lowest energy state is necessary, it is useful to first sort the results to identify the multiplicity for each value and avoid redundant computation. 
\par
These three stages represent parameter models for the split-execution application. In particular, stages 1 and 3 occur within the conventional RAM model while stage 2 is executed within the adiabatic quantum computing model. The translation between these two models is signified by the map of the logical Hamiltonian to the physical hardware, i.e., minor embedding.
\begin{figure}
\begin{lstlisting}
model Stage3
{
  param LPS = 0
  param Success   = 0.75
  param Accuracy  = 0.99
  param Results   = ceil(log(1-(Accuracy))/log(1-Success))                                                                                        
  param Length = LPS
  param SortOps = log(Results) * Results 

  data R as Array(Results, LPS)

  kernel FindSolution {
          execute  sort [1] {
      loads [Results] of size [4*Length]
      flops [SortOps] as sp
      stores [Results] to R
    }
  }

  kernel main {
          FindSolution
  }
}
\end{lstlisting}
\caption{Stage 3 of the split-execution application parses the readout results and sorts the solutions to recover the optimization result.}
\label{fig:stage3}
\end{figure}
\subsection{Timing Results}
\label{sec:results}
There are several key characteristics that must be measured when analyzing the split-execution 
performance. We have chosen to focus on those highlighted metrics, namely, performance and scaling. 
In particular, we measure performance in terms of the total execution time required by the system 
to return a solution. This includes both the programming and execution steps of the application. 
We also focus on scaling as a measure of how the time-to-solution varies with the size of the problem.
\begin{figure}
\textbf{(a)}\hspace{-1cm}
\begin{subfigure}{0.5\textwidth}
\centerline{\includegraphics[height=4cm]{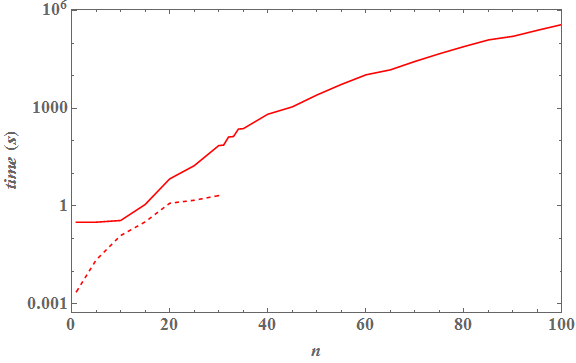}}
\end{subfigure}
\textbf{(b)}\hspace{-1cm}
\begin{subfigure}{0.5\textwidth}\vspace{1em}
\centerline{\includegraphics[height=4cm]{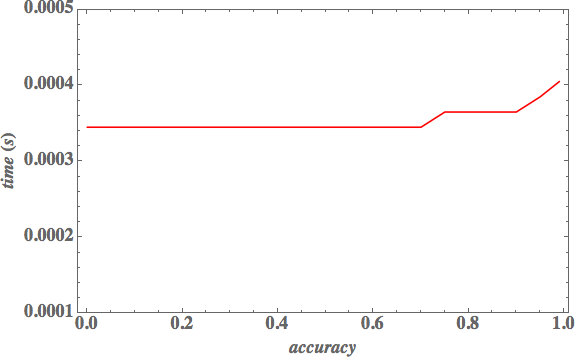}}
\end{subfigure}
\textbf{(c)}\hspace{-1cm}
\begin{subfigure}{0.5\textwidth}\vspace{1em}
\centerline{\includegraphics[height=4cm]{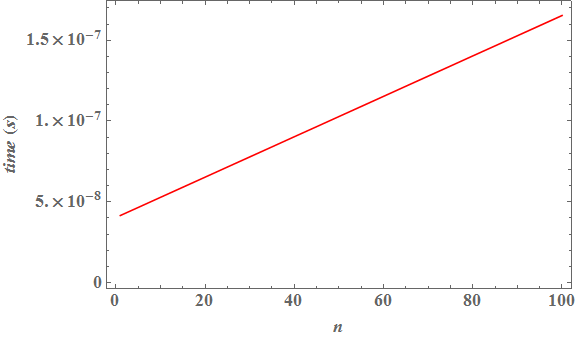}}
\end{subfigure}
\caption{The performance scaling for the 3 stages of the split-execution program. (a) Stage 1 timing in seconds with respect to input problem size $n = \texttt{LPS}$ from ASPEN model (solid line) and experimentally observed timing (dashed line); (b) Stage 2 timing with respect to desired accuracy; (c) Stage 3 timing with respect to input size.}
\label{fig:stage_scaling}
\end{figure}
\par
The measure of what dominates the performance cost is found by looking at the timing plots for each 
stage of the application model. These are shown in Fig.~\ref{fig:stage_scaling}(a)-(c). In Fig.~\ref{fig:stage_scaling}(a), the stage 1 timing is shown on a logarithmic scale with respect to the size $n$ of a complete input graph. The solid line represents the results from the ASPEN model over the range of 1 to 100, while the dashed line represents experimentally measured timings for the Cai, Macready, and Roy algorithm taken from Ref. \cite{Cai2014} for inputs from 1 to 30. Over the latter region, the ASPEN models is within a factor of 4 of the experimentally observed timing except in the region $n <10$, which it overestimates. Our analysis has assumed the worst-case behavior for the embedding, whereas Cai et al.~reported that the average case behavior scales appears to scale linearly for a fixed hardware graph. 
\par
By comparison, Fig.~\ref{fig:stage_scaling}(b) shows the timing for stage 2 in the application model with respect to the desired accuracy $p_a$ in the computed solution. The plotted results correspond to a value of $p_s = 0.7$. However, we have found that this performance curve is approximately the same for all values of $p_s > 0.6$. In each  of these instances, the number of iterations necessary to realize over an accuracy $p_a > 0.99$ is so few that the overall timing for stage 2 is well below the time of stage 1. Finally, the stage 3 performance is presented in Fig.~\ref{fig:stage_scaling}(c) as the timing of the linear sort routine with respect to the input problem size. The nearly linear dependence provides a very small contribution to the overall timing for the program.
\par
It is apparent from these results that the stage 1 timing increases dramatically with input graph size and dominates the costs for this application model. The relatively steep increase in timing traces back to the graph minor embedding calculation that is required to program the D-Wave processor. Whereas our analysis has assumed the worst-case behavior for the probabilistic algorithm from Cai et al., the stark discrepancy in the timings between stages would seem to be true for any embedding algorithm that scales with problem size. The significance of this conclusion is that the application bottleneck does not lie in quantum program execution (stage 2) but rather in the purely classical pre-processing state. This computational complexity expresses the difficulty in translation from the conventional QUBO problem instance into the hardware constrained quantum program.
\par
The three-staged application model described here is very common for early descriptions of the using the D-Wave processor. However, as these results indicate, the bottleneck that arises from inlining the embedding calculation with the run-time environment is a consequence of a poor programming decision. Rather it may be beneficial to use some variant of off-line embedding, in which specific input graphs are pre-embedded and stored in a graph lookup table. The difficulty of these off-line calculations would not impact the performance of the application model, but use of the lookup table would require some variant of graph isomorphism to identify which embedding to apply. The graph isomorphism problem has recently been shown to be solvable using adiabatic quantum computing \cite{Gaitan2014,Zick2015}, raising the prospects the D-Wave processor could be used to program the D-Wave processor!
\section{Conclusion}
\label{sec:con}
We have presented an executable model of a split-execution computing environment. We have defined both machine and programming models for a mixed computation that offloads an optimization task to a QPU. Our approach has been based on using the ASPEN modeling language to develop representations of the CPU and QPU hardware as well as a generic application that makes use of both quantum and classical resource. We have also presented an analysis of the performance and scaling of this program on the devised architecture and we have shown that the embedding step grows quickly as the size of the input graph increases linearly. Compared to the execution time needed to collect sufficient samples from the QPU, these results indicate that the pre-processing required for the application greatly exceeds quantum execution time. Ultimately, these pre-processing costs trace back to the minor graph embedding problem.
\par
We believe these results suggest two important aspects of split-execution computing. First, the  pre-processing overhead for split-execution must be reduced by many orders of magnitude in order to become processor limited. It is not immediately clear if this is possible, but research into fast minor embedding will be essential for this purpose. However, it must also be considered that our models have not exploited more sophisticated host systems, e.g., HPC, or more sophisticated algorithms, and there may be additional parallel strategies that can accelerate the pre-processing stage.
\section*{Acknowledgments}
This work was supported by the United States Department of Defense (DoD) and used resources of the Computational Research and Development Programs at Oak Ridge National Laboratory. This manuscript has been authored by UT-Battelle, LLC, under Contract No.~DE-AC0500OR22725 with the U.S. Department of Energy.

\bibliographystyle{IEEEtran}
\bibliography{apdcm2016}

\end{document}